\documentstyle[aps,prl,twocolumn,epsf]{revtex}

\begin{document}
\draft

\title{
Temperature-dependent relaxation times in a
trapped Bose-condensed gas}

\author{T. Nikuni and A. Griffin}

\address{
Department of Physics, University of Toronto, Toronto, Ontario M5S 1A7, Canada}

\date{\today}

\maketitle

\begin{abstract}
Explicit expressions for all the transport coefficients have recently been
found for a trapped Bose condensed gas at finite temperatures.
These transport coefficients are used to define the characteristic relaxation times,
which determine the crossover between the mean-field collisionless and the
two-fluid hydrodynamic regime.
These relaxation times are evaluated
as a function of the position in the trap potential.
We show that all the relaxation times are dominated by the collisions
between the condensate and the non-condensate atoms, and are much smaller
than the standard classical collision time used in most of the current literature.
The 1998 MIT study of the collective modes at finite temperature is
shown to have been well within the two-fluid hydrodynamic regime.
\end{abstract}

\pacs{PACS numbers: 03.75.Fi, 05.30.Jp, 67.40.Db }

Recently, the complete set of two-fluid hydrodynamic equations were derived starting
from a microscopic model for a
trapped Bose-condensed gas at finite temperatures \cite{LK}.
These included all the transport processes which arise in a Bose superfluid, namely
the thermal conductivity, shear viscosity, and the four coefficients of second
viscosity.
Working within the Thmas-Fermi approximation,
these two-fluid hydrodynamic equations can be written in the well-known
Landau-Khalatnikov form \cite{Landau,Khal} used to describe the hydrodynamic behavior of
superfluid $^4$He.
In the present paper, we use the explicit expressions for the transport coefficients
obtained in Ref.~\cite{LK} to define three characteristic relaxation times
associated with distinct relaxation processes due to collisions between atoms.
As with a classical gas, relaxation times which are associated with 
transport coefficients
play a crucial role in understanding the dynamics of a Bose-condensed gas.
For example, they determine how fast thermal equilibrium is reached when the gas is
perturbed, the rate of evaporative cooling in the Bose-condensed phase, and the
characteristic frequency dividing the collisionless from the collision-dominated
hydrodynamic region.

We recall that the transport coefficients describe the
deviations of the non-condensate distribution function
$f({\bf r},{\bf p},t)$ from the ``local thermal equilibrium'' form \cite{huang}.
Thus it is natural that they describe the rate at which such deviations
relax.
Ref.~\cite{LK} gives all the various transport coefficients explicitly as
functions of the local fugacity, or, effectively, the local static 
condensate density $n_{c0}({\bf r})$.
However these expressions are quite complex and involve various
integrals which must be done numerically.
In the present paper, we evaluate the associated transport relaxation times as
a function of position in the trap, for experimental parameters which have been
used in recent experiments.
In particular, we show that the pioneering MIT study \cite{MIT} of collective modes at finite
temperatures was well within the hydrodynamic region, contrary to what was originally
estimated in that paper.

In the absence of a quantitative theory of the collision times in a
Bose-condensed gas, most of the BEC literature up to now has used the
Maxwell-Boltzmann elastic collision time \cite{huang,NZG}
\begin{equation}
\frac{1}{\tau_{\rm cl}({\bf r})}=\sqrt{2}\tilde n_0({\bf r})\sigma \bar{v},
\label{tau_cl}
\end{equation}
as an estimate of the relaxation rates in trapped Bose gases.
In particular, $\omega \tau_{\rm cl}=1$ has been used \cite{MIT} to estimate the
``crossover'' between collisionless modes ($\omega\tau_{\rm cl}\gg 1$) and the
collision-dominated hydrodynamic modes ($\omega\tau_{\rm cl}\ll 1$).
In (\ref{tau_cl}), $\tilde n_0({\bf r})$ is the local density of the thermally
excited atoms, $\sigma=8\pi a^2$ is the Bose atomic cross-section in the $s$-wave
approximation.
$\bar{v}=\sqrt{ 8k_{\rm B}T/\pi m}$
is the mean value of the velocity of the thermal atoms in a classical gas
(which is very close to that in an ideal Bose gas).
In the detailed solution of the Boltzmann equation for $f({\bf r},{\bf p},t)$
for a trapped Bose condensed gas at finite temperatures \cite{LK,NZG,ZNG},
an important distinction is made between collisions
between thermal atoms ($C_{22}$ collisions) and the collision integral involving
atoms being scattered into/out the condensate ($C_{12}$ collisions).
In the classical limit, the $C_{22}$ collisions give rise to the collision time
given in (\ref{tau_cl}).
$C_{12}$ collisions only appear at temperatures below the Bose-Einstein
transition temperature $T_{\rm BEC}$ and are thus a
unique feature of a Bose-condensed gas.
Our detailed calculations in Ref.~\cite{LK} and in this paper include
both the $C_{22}$ and $C_{12}$ collisions.
However, we find that the various transport
relaxation times for the thermal atoms are dominated by the $C_{12}$ collisions in
a trapped Bose gas because of the high condensate density.
A key feature about trapped gases is that the condensate is strongly peaked
(high density) at the center of the trap while the thermal cloud (non-condensate
distribution) is always spread out over a much larger region.
Even at $T=0.99T_{\rm BEC}$, for example, the condensate density at the center of the
trap is still comparable to the non-condensate density.

Our major finding in this paper is that the relaxation time $\tau_{\kappa}$
associated with the thermal conductivity is well approximated by a simple
formula
\begin{equation}
\frac{1}{\tau_{\kappa}({\bf r})}\simeq \sqrt{2}n_{c0}({\bf r})\sigma \bar{v}
\equiv \frac{1}{\tau_{\rm BE}({\bf r})},
\label{tau_BE}
\end{equation}
as long as
\begin{equation}
n_{c0}({\bf r})\gg \tilde n_0({\bf r}).
\label{density}
\end{equation}
Moreover, numerical results show that all three characteristic relaxation
times ($\tau_{\mu},\tau_{\kappa}$ and $\tau_{\eta}$) are of the same magnitude,
within a factor of two.
Thus we conclude that to a good first approximation, the relaxation time 
$\tau_{\rm BE}({\bf r})$ as defined
in (\ref{tau_BE}) can be used as a simple ``universal'' collision time in trapped
Bose gases, replacing the expression in (\ref{tau_cl}).
One immediately sees that since (\ref{density}) is valid
in the center of the trap at effectively all temperatures below $T_{\rm BEC}$,
we obtain $\tau_{\rm BE}({\bf r})\ll\tau_{\rm cl}({\bf r})$.
For modes concentrated in the condensate core region,
it is $\tau_{\rm BE}({\bf r})$ which determines the crossover frequency between
the collisionless and hydrodynamic frequency regions.
In the thermal gas outside the condensate core, $\tau_{\rm cl}$ in
(\ref{tau_cl}) gives a good estimate for the transport relaxation times.

We now turn to the detailed calculations which lead to the above conclusions.
Referring to Appendix B of Ref.~\cite{LK}, we first summarize the formal expressions
for the various relaxation times in a trapped Bose gas:

\bigskip

{\it 1. Thermal conductivity} :
\begin{equation}
\kappa({\bf r})=\frac{5}{2}\tau_{\kappa}({\bf r})
\frac{\tilde n_0({\bf r})k_{\rm B}^2T}{m}
\left\{ \frac{7g_{7/2}(z_0)}{2g_{3/2}(z_0)}-\frac{5}{2}\left[\frac{g_{5/2}(z_0)}{g_{3/2}(z_0)}
\right]^2\right\},
\label{kappa}
\end{equation}
where $z_0\equiv e^{\beta [\mu_{c0}-U_0({\bf r})]}$ is the local fugacity
and $g_n(z)$ are the usual Bose-Einstein functions \cite{erata}.
The effective potential
$U_0({\bf r})\equiv U_{\rm ext}({\bf r})+2g[n_{c0}({\bf r})+\tilde n_0({\bf r})]$
includes the external trap potential $U_{\rm ext}({\bf r})$ as well as the
self-consistent Hartree-Fock mean field. 
The interaction between atoms 
is treated by the usual low energy $s$-wave scattering length approximation,
with $g=4\pi \hbar^2 a/m$.
The thermal relaxation time $\tau_{\kappa}$ in (\ref{kappa}) is given by
\begin{equation}
\frac{1}{\tau_{\kappa}}=\frac{1}{\tau_{\kappa,12}}+\frac{1}{\tau_{\kappa,22}},
\label{tau_kappa}
\end{equation}
with
\begin{equation}
\tau_{\kappa,12}=
\tau_{\rm BE}\frac{15\sqrt{2}\pi^{7/2}}{4}
\frac{\frac{7}{2}g_{7/2}(z_0)g_{3/2}(z_0)-\frac{5}{2}g_{5/2}^2(z_0)}
{\tilde n_0\Lambda_0^3 I_{12}^{\kappa}(z_0)},
\label{tau_kappa12}
\end{equation}
\begin{equation}
\tau_{\kappa,22}=
\tau_{\rm cl}\frac{15\sqrt{2}\pi^{7/2}}{4}
\frac{\frac{7}{2}g_{7/2}(z_0)g_{3/2}(z_0)-\frac{5}{2}g_{5/2}^2(z_0)}
{I_{22}^{\kappa}(z_0)},
\label{tau_kappa22}
\end{equation}
where $\tau_{\rm cl}$ and $\tau_{\rm BE}$ are defined in (\ref{tau_cl}) and
(\ref{tau_BE}) respectively.
The equilibrium densities of the condensate
and non-condensate are given by
$\tilde n_0({\bf r})=(1/\Lambda_0^3)g_{3/2}(z_0)$ and
$n_{c0}({\bf r})=[\mu_{c0}-U_{\rm ext}({\bf r})]/g-2\tilde n_0({\bf r})$,
where $\Lambda_0\equiv(2\pi\hbar^2/mk_{\rm B}T)^{1/2}$ is the thermal
de Broglie wavelength.
The dimensionless integrals $I_{22}^{\kappa}$ and $I_{12}^{\kappa}$ are defined in
Ref.~\cite{LK} and involve an integration over the $C_{12}$ and $C_{22}$ collision
cross-sections (which involve products of the
equilibrium Bose distributions describing the thermal atoms
with an appropriate weighting function).

\bigskip

{\it 2. Shear viscosity} :
\begin{equation}
\eta({\bf r})
=\tau_{\eta}({\bf r})\tilde n_0({\bf r})k_{\rm B}T\left[\frac{g_{5/2}(z_0)}{g_{3/2}(z_0)}\right],
\label{eta}
\end{equation}
where the viscous relaxation time $\tau_{\eta}$ is given by
\begin{equation}
\frac{1}{\tau_{\eta}}\equiv\frac{1}{\tau_{\eta,12}}+\frac{1}{\tau_{\eta,22}},
\label{tau_eta}
\end{equation}
with
\begin{equation}
\tau_{\eta,12}
=\tau_{\rm BE}\frac{5\sqrt{2}\pi^{7/2}}{2}
\left[\frac{g_{5/2}(z_0)g_{3/2}(z_0)}{\tilde n_0\Lambda_0^3 I_{12}^{\eta}(z_0)}\right],
\label{tau_eta12}
\end{equation}
\begin{equation}
\tau_{\eta,22}
=\tau_{\rm cl}\frac{5\sqrt{2}\pi^{7/2}}{2}
\left[\frac{g_{5/2}(z_0)g_{3/2}(z_0)}{I_{22}^{\eta}(z_0)}\right].
\label{tau_eta22}
\end{equation}
Again, the dimensionless integrals $I_{12}^{\eta}$ and $I_{22}^{\eta}$ are
given by somewhat complicated expressions involving products of the equilibrium Bose 
distribution functions \cite{LK}.

\bigskip

{\it 3. Coefficients of second viscosity} :

From general considerations, Landau and Khalatnikov showed \cite{Khal} that in a 
Bose superfluid, there are four coefficients of second viscosity.
We find in our model that \cite{LK}
\begin{eqnarray}
\zeta_1&=&\zeta_4=\frac{gn_{c0}({\bf r})}{3m}\sigma_H\tau_{\mu}, \cr
\zeta_2&=&\frac{gn_{c0}^2({\bf r})}{9}\sigma_H\tau_{\mu},~
\zeta_3=\frac{g}{m^2}\sigma_H\tau_{\mu},
\label{zeta}
\end{eqnarray}
where the relaxation time $\tau_{\mu}$ describes how fast the chemical potentials
of the condensate and non-condensate components come to equilibrium.

The relaxation time $\tau_{\mu}$, which
arises entirely from the $C_{12}$ collision integral in the kinetic equation,
was first discussed in detail in Refs.~\cite{NZG,ZNG}.
The hydrodynamic renormalization factor $\sigma_H$ in (\ref{zeta})
only involves the static
thermodynamic functions of the gas, and is defined in Refs.~\cite{NZG,ZNG}.
More explicitly, one finds
\begin{equation}
\frac{1}{\tau_{\mu}}=\left(\frac{gn_{c0}}{k_{\rm B}T}\right)\frac{1}{\tau_{12}\sigma_H},
\label{tau_12}
\end{equation}
where $\tau_{12}$ is a collision time associated with the $C_{12}$ collisions
\cite{NZG,ZNG}.
We note that in a Bose-condensed gas, the coefficients of second
viscosity arise entirely from the $C_{12}$ collisions, in contrast to the 
$\kappa$ and $\eta$
transport coefficients, which also have a contribution from $C_{22}$ collisions
(which, however, is very small in a trapped gas).

\begin{figure}[ht]
 \epsfxsize=80mm
  \centerline{\epsfbox{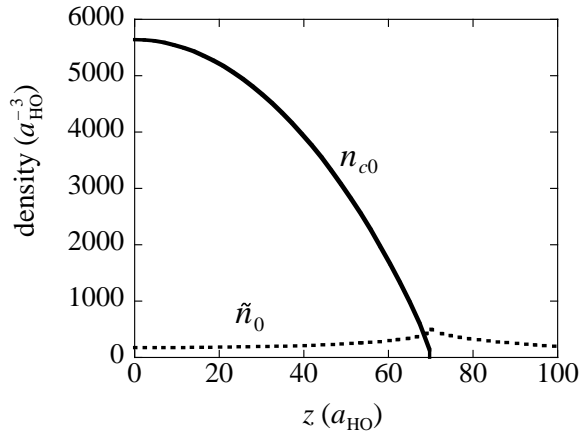}}
\caption{The Thomas-Fermi density profiles along the $z$ axis of the condensate
and the non-condensate at $T=1\mu$K ($T\simeq 0.7T_{\rm BEC}$) with
$N=4\times 10^7$ atoms.}
\label{fig:density}
\end{figure}

From Eqs.~(\ref{tau_kappa12}) and (\ref{tau_kappa22}), one sees that
$\tau_{\rm BE}/\tau_{\kappa,12}$ and $\tau_{\rm cl}/\tau_{\kappa,22}$ are
given as universal functions of the local fugacity $z_0({\bf r})$.
It turns out that $\tau_{\rm BE}/\tau_{\kappa,12}\simeq 1$ for 
$z_0\gtrsim 0.3$ and $\tau_{\rm cl}/\tau_{\kappa,22}\approx 8/15$ for
$0\leq z_0 \leq 1$.
Similarly, we see from Eqs.~(\ref{tau_eta12}) and (\ref{tau_eta22})
that $\tau_{\rm BEC}/\tau_{\eta,12}$ and $\tau_{\rm cl}/\tau_{\eta,22}$ are
also universal functions of $z_0$, but now have a more complicated dependence,
increasing as $z_0$ increase.
These analytical results are confirmed in Figs.~2 and 3.
We emphasize that while $\tau_{\rm BE}$ in (\ref{tau_BE}) gives a good estimate
of $\tau_{\kappa,12}$ in (\ref{tau_kappa12}), it is an empirical formula.

We note that the transport relaxation rates involve integrations
which suppress the weight of the low energy collisions.
As a result, there is no Bose enhancement from these processes and no 
divergence in $1/\tau_{\kappa,\eta}$ when $z_0=e^{-\beta g n_{c0}({\bf r})}$
approaches unity (at the edge of the condensate or at $T_{\rm BEC}$).
This is in contrast with the collision times $\tau_{12}$ and $\tau_{22}$
defined as in Refs.~\cite{NZG,ZNG}, in which $1/\tau_{12}$ and $1/\tau_{22}$
have an infrared divergence for $z_0=1$ due to the contribution from
low momentum collisions.
However, note that $1/\tau_{\mu}$ in (\ref{tau_12}) has no divergence at
$z_0=1$ \cite{NZG}.

As a concrete illustration of the behavior of these various transport relaxation times,
we consider
the MIT study of collective modes at finite temperatures \cite{MIT}.
The trap frequencies were $\omega_x/2\pi=\omega_y/2\pi=230$Hz and
$\omega_z/2\pi=18$Hz, the scattering length $a=2.75$nm for $^{23}$Na,
and the total number of atoms was $N=4\times 10^7$.
Both the $m=0$ quadrupole oscillations and the
out-of-phase dipole oscillations were studied in Ref.~\cite{MIT}.
In both cases, the condensate and the thermal cloud oscillate along the
$z$-axis with the collective mode frequency of the order of the axial trap frequency
$\omega_z$.
It is therefore convenient to compare the various transport relaxation rates with the
trap frequency $\omega_z$.

\begin{figure}[ht]
 \epsfxsize=80mm
  \centerline{\epsfbox{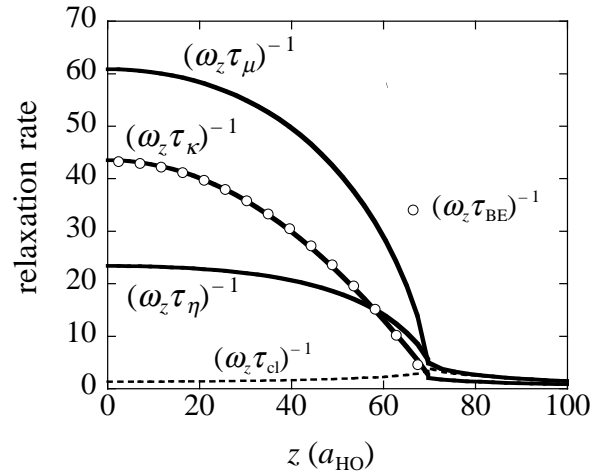}}
\caption{
The position dependence of the various dimensionless relaxation rates
at $T=1\mu$K ($T\simeq 0.7T_{\rm BEC}$)with $N=4\times 10^7$.
The open circles show the approximate relaxation rate
$1/\omega_z\tau_{\rm BE}$ using Eq.~(\ref{tau_BE}) and the dashed
line is the classical collision rate $1/\omega_z\tau_{\rm cl}$ using
Eq.~(\ref{tau_cl}).}
\label{fig:tau_1}
\end{figure}

In Fig.~1, we plot the spatial dependence of both the condensate and
non-condensate densities along the axial $z$-direction at 
$T\simeq 0.7T_{\rm BEC}$.
In Fig.~2, we plot the spatial dependence of the
various relaxation rates.
We find that all the relaxation rates are much larger than the
trap frequency $\omega_z$, and thus the collective oscillations with
the frequency $\omega\sim\omega_z$ are hydrodynamic modes.
In Fig.~2, we also show the approximate relaxation rates using
the simple formula given by (\ref{tau_BE}).
This empirical expression is seen to be in excellent agreement with the
thermal relaxation time $\tau_{\kappa}$.
For comparison, we also plot the classical elastic collision time
given by (\ref{tau_cl}).
Clearly this classical collision rate is much smaller than all the
transport relaxation rates due to $C_{12}$ collisions.
Outside the central condensate region, $\tau_{\mu}^{-1}$ vanishes and
$\tau_{\kappa}^{-1}$ and $\tau_{\eta}^{-1}$ only have contributions from the
$C_{22}$ collisions.
They can be well approximated by the classical gas results
$\tau_{\kappa}\simeq (15/8)\tau_{\rm cl}$ and
$\tau_{\eta}\simeq (5/4)\tau_{\rm cl}$.
To see the dependence on the total number of the atoms, we plot in Fig.~3
the relaxation times for $N=10^5$.
In this case, all the relaxation rates are such that $\omega_z\tau \sim 1$
and thus the collective modes would be in the transition region between
the collisionless and hydrodynamic domains.

We might also note that the diffusive equilibrium relaxation time $\tau_{\mu}$
is always slightly smaller than the other relaxation times (see Figs.~2 and 3).
As we have discussed in Ref.~\cite{NZG,ZNG}, in principle one could have
$\omega\tau_{\mu}\gg 1$ even in the hydrodynamic region where
$\omega\tau_{\kappa},\omega\tau_{\eta} \ll 1$.
In a uniform Bose gas, this can occur in a fairy wide region near $T_{\rm BEC}$ 
where $\tau_{\mu}$
becomes very large \cite{NZG}.
In this region, one expects to find a new zero-frequency relaxational mode
\cite{NZG,CJP}.
However, the high condensate density in the center of a trap (relative to the
low density of the thermal cloud) precludes one from observing this
diffusive relaxational mode - unless one works extremely 
close to $T_{\rm BEC}$.

\begin{figure}[ht]
\epsfxsize=80mm
\centerline{\epsfbox{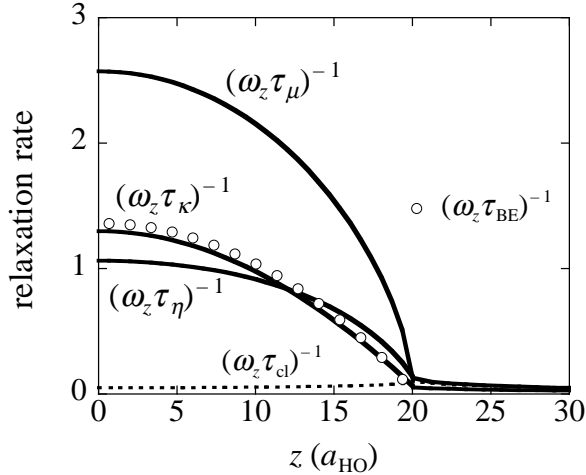}}
\caption{
Plot of the dimensionless relaxation rates for $N=10^5$ at $T=160n$K ($T\simeq
0.8 T_{\rm BEC}$).}
\label{fig:taus}
\end{figure}

One can evaluate the peak value of the approximate
relaxation rate in (\ref{tau_BE}) using
the $T=0$ Thomas-Fermi expression for the peak density
$n_{c0}(0)$ at the center of the trap.
This gives the useful formula:
\begin{equation}
\frac{1}{\bar\omega\tau_{\rm BE}(0)}\simeq
6.47\left(\frac{a}{a_{\rm HO}}\right)^{7/5}N^{17/30}
\left(\frac{T}{T_c^0}\right)^{1/2}
\left(\frac{N_c}{N}\right)^{2/5},
\label{tau_approx}
\end{equation}
where $a_{\rm HO}\equiv (\hbar/m\bar\omega)^{1/2}$, 
$\bar\omega\equiv (\omega_x\omega_y\omega_z)^{1/3}$ is the averaged
frequency of the anisotropic trap, $N_c$ is the total number of the
atoms in the condensate,
and $T_c^0\equiv (\hbar\bar\omega/k_{\rm B})
[N/\zeta(3)]^{1/3}$ is the transition
temperature of an ideal trapped Bose gas.
For a given value of $N$, this expression can be used to estimate at what
temperature we pass from the collisionless to the hydrodynamic region for
a given trap ($\omega\tau_{\rm BE}(0)=1$).


To be in the two-fluid hydrodynamic region ($\omega\tau_{\rm BE}<1$),
one must work with a sample size with $N\sim 10^6$ atoms or more, when
dealing with $^{87}$Rb or $^{23}$Na atoms.
One can also make use of a highly anisotropic trap potential
$\omega_x,\omega_y\gg\omega_z$ to have low-frequency collective mode
$\omega\sim\omega_z$ while keeping the high density due to the
tight confinement in the radial direction.
We note that if one is already close to the hydrodynamic domain above $T_{\rm BEC}$
as estimated by using (\ref{tau_cl}) \cite{MIT,ENS}, one will be deeply in the
hydrodynamic region $(\omega\tau_{\rm BE}\ll 1)$
when the Bose-condensate forms.
This is simply a result of the fact that in the condensate core region,
$\tau_{\rm BE}\ll\tau_{\rm cl}$.

The scattering length of metastable $^4$He$^*$ \cite{ENS} is estimated to be
$a\sim 16$nm, much larger than in $^{87}$Rb or $^{23}$Na.
Combined with the large sample size ($N\sim 10^7$), the recent Bose condensates
observed in $^4$He$^*$ are well within the hydrodynamic domain even above $T_{\rm BEC}$.
Finally, the adjustable value of $a$ achievable near a Feshbach resonance
(see Ref.~\cite{JILA}) could be used to study collective mode frequencies as
one smoothly goes from the mean-field (small $a$)
to the collision-dominated hydrodynamic (large $a$) region.

Detailed predictions for the collective mode frequencies in the two-fluid
hydrodynamic region of superfluid trapped gases will be presented elsewhere,
based on the theory given in Ref.~\cite{LK}.
In trapped gases, one has the interesting situation that while
the thermal cloud in the condensate
region may be well in the hydrodynamic region, one eventually always
enters the collisionless region in the low density tail of the thermal cloud.
This has not been included in the derivation of the two-fluid hydrodynamic
equations given in Ref.~\cite{LK}.
While this has little effect on the frequency of a condensate mode,
it may be important in the damping of certain modes.
Such effects have been included above $T_{\rm BEC}$ by using a cutoff
\cite{KPS,NG}.
The main point, however, is that the damping in the hydrodynamic region will
be directly proportional to an appropriate spatial average of
the transport relaxation times introduced in this paper.

We thank Eugene Zaremba for extensive discussions as well as comments on the
manuscript.
T.N. is supported by JSPS of Japan and A.G. is supported by NSERC of Canada.



\begin{references}
\bibitem{erata} Eq.~(125) of Ref.~\cite{LK} has some typos.
The corrected result is given here.
\end{references}
\end{document}